\documentclass[twocolumn,english,aps,prl,superscriptaddress,preprintnumbers]{revtex4}
\usepackage{lmodern}
\usepackage[T1]{fontenc}
\usepackage[latin9]{inputenc}
\usepackage{verbatim}
\usepackage{amsmath}
\usepackage{amssymb}
\usepackage{graphicx}
\usepackage{setspace}
\usepackage{esint}
\makeatletter
\@ifundefined{textcolor}{}
{%
 \definecolor{BLACK}{gray}{0}
 \definecolor{WHITE}{gray}{1}
 \definecolor{RED}{rgb}{1,0,0}
 \definecolor{GREEN}{rgb}{0,1,0}
 \definecolor{BLUE}{rgb}{0,0,1}
 \definecolor{CYAN}{cmyk}{1,0,0,0}
 \definecolor{MAGENTA}{cmyk}{0,1,0,0}
 \definecolor{YELLOW}{cmyk}{0,0,1,0}
 }

\usepackage{units}\usepackage{amsfonts}

\usepackage[T1]{fontenc}
\usepackage{lmodern}

\newcommand{\meff}{\ensuremath{m_\mathrm{eff}}}
\newcommand{\kB}{\ensuremath{k_\mathrm{B}}}
\newcommand{\Gm}{\ensuremath{\Gamma_\mathrm{m}}}
\newcommand{\Geff}{\ensuremath{\Gamma_\mathrm{eff}}}
\newcommand{\Gdba}{\ensuremath{\Gamma_\mathrm{dba}}}
\newcommand{\xzp}{\ensuremath{x_\mathrm{zpf}}}

\newcommand{\Om}{\ensuremath{\Omega_\mathrm{m}}}
\newcommand{\Omod}{\ensuremath{\Omega_\mathrm{mod}}}

\newcommand{\ocavity}{\ensuremath{\omega_\mathrm{c}}}
\newcommand{\opump}{\ensuremath{\omega_\mathrm{pump}}}
\newcommand{\oprobe}{\ensuremath{\omega_\mathrm{probe}}}

\newcommand{\nm}{\ensuremath{\bar{n}_\mathrm{m}}}
\newcommand{\np}{\ensuremath{\bar{n}_\mathrm{p}}}

\newcommand{\ahat}{\ensuremath{\hat{a}}}
\newcommand{\xhat}{\ensuremath{\hat{x}}}

\newcommand{\phat}{\ensuremath{\hat{p}}}

\newcommand{\Hint}{\ensuremath{\hat{H}_\mathrm{int}}}

\usepackage{babel}

\makeatother

\usepackage{babel}
\begin{document}

\title{Control of microwave signals using circuit nano-electromechanics}

\author{X.~Zhou}

\affiliation{\'{E}cole Polytechnique F\'{e}d\'{e}rale de Lausanne (EPFL), 1015, Lausanne,
Switzerland}

\affiliation{Max-Planck-Institut f\"{u}r Quantenoptik, 85748 Garching, Germany}

\author{F.~Hocke}

\affiliation{Walther-Meissner-Institut, 85748 Garching, Germany}

\author{A.~Schliesser}

\affiliation{\'{E}cole Polytechnique F\'{e}d\'{e}rale de Lausanne (EPFL), 1015, Lausanne,
Switzerland}

\affiliation{Max-Planck-Institut f\"{u}r Quantenoptik, 85748 Garching, Germany}

\author{A.~Marx}

\affiliation{Walther-Meissner-Institut, 85748 Garching, Germany}

\author{H.~Huebl}

\email{hans.huebl@wmi.badw.de}

\affiliation{Walther-Meissner-Institut, 85748 Garching, Germany}

\author{R.~Gross}

\affiliation{Walther-Meissner-Institut, 85748 Garching, Germany}

\affiliation{Technische Universit\"{a}t M\"{u}nchen, 85748 Garching, Germany}

\author{T.~J.~Kippenberg}

\email{tobias.kippenberg@epfl.ch}

\affiliation{\'{E}cole Polytechnique F\'{e}d\'{e}rale de Lausanne (EPFL), 1015, Lausanne,
Switzerland}

\affiliation{Max-Planck-Institut f\"{u}r Quantenoptik, 85748 Garching, Germany}

\begin{abstract}
Microwave superconducting coplanar waveguide resonators are
crucial elements in sensitive astrophysical detectors \cite{day_broadband_2003}
and circuit quantum electrodynamics (cQED) \cite{Wallraff2004}.
Coupled to artificial atoms in the form of superconducting qubits
\cite{vion_manipulating_2002,Clarke2008}, they now provide
a technologically promising and scalable platform for quantum information
processing tasks\cite{Wallraff2004,Houck2007,Majer2007,Schoelkopf2008,Mariantoni2011}.
Coupling these circuits, \textit{in situ}, to other
quantum systems, such as molecules \cite{Rabl2006,Andre2006},
spin ensembles \cite{Schuster2010,Kubo2011}, quantum dots
\cite{frey_dipole_2012}  or mechanical oscillators \cite{Regal2008,Rocheleau2010,Marquardt2009,Kippenberg2008}
has been explored to realize hybrid systems with extended functionality.
Here, we couple a superconducting coplanar waveguide resonator to
a nano-mechanical oscillator, and demonstrate all-microwave
field controlled slowing, advancing and switching of microwave signals.
This is enabled by utilizing electromechanically induced transparency
\cite{Weis2010,Safavi-Naeini2011,Teufel2011}, an effect
analogous to electromagnetically induced transparency (EIT) in atomic
physics  \cite{Fleischhauer2005}. The exquisite temporal
control gained over this phenomenon provides a route towards realizing
advanced protocols for storage of both classical and quantum microwave
signals \cite{Clerk2012,Braunstein2003,Tian2012}, extending the toolbox
of quantum control techniques of the microwave field. 

\end{abstract}

\maketitle

Cavity opto- and electro-mechanical systems \cite{Marquardt2009,Kippenberg2008}
realize parametric coupling of an electromagnetic resonance to a mechanical
mode, which enables a wide range of phenomena including displacement
measurements, sideband cooling or amplification of mechanical motion.
In the microwave domain, this physics has been explored by coupling
of superconducting cavities to micro- and nano-mechanical oscillators
\cite{Regal2008,Teufel2011,Rocheleau2010} enabling efficient transduction
of mechanical motion with an imprecision below the level of the zero
point motion \cite{Teufel2009}, electro-mechanical sideband cooling
to the quantum ground state \cite{Teufel2011b,Rocheleau2010}, and
electromechanical amplification of microwave signals \cite{massel_microwave_2011}.
Moreover the mutual coupling of microwaves and mechanical oscillator
modifies the microwave response leading to the phenomenon of electromechanically
induced transparency \cite{Weis2010}. Here we demonstrate that superconducting
circuit nano-electromechanical systems enable a novel class of control
phenomena over the microwave field. Exploiting the coupling of a superconducting
microwave cavity to a nano-mechanical oscillator, we demonstrate tunable,
sub- and superluminal microwave pulse propagation, mediated by the
nano-mechanical oscillator's response. The high mechanical quality
factor enables a maximum delay of microwave pulses exceeding 3 ms,
corresponding to an effective coaxial cable length of several hundreds
of kilometers. Importantly, this delay is achieved with negligible
losses and pulse distortion. Moreover, we explore the circuit nano-electromechanical
response to a time-dependent control field, which is required for
a series of advanced protocols including quantum state transfer and
storage \cite{Verhagen2011,Romero-Isart2011}, fast sideband cooling
\cite{Wang2011a}, as well as switching, modulation and routing of
classical and quantum microwave signals. To this end, we demonstrate
all-microwave field controlled switching and show that counterintuitive
regimes can be found in which the switching time can be much faster
than the mechanical oscillator energy decay time. Finally, we also
demonstrate mapping of the mechanical$\,$(Duffing) nonlinearity into
the microwave domain. 

\begin{figure}[tbp]
\centering \includegraphics[scale=0.4]{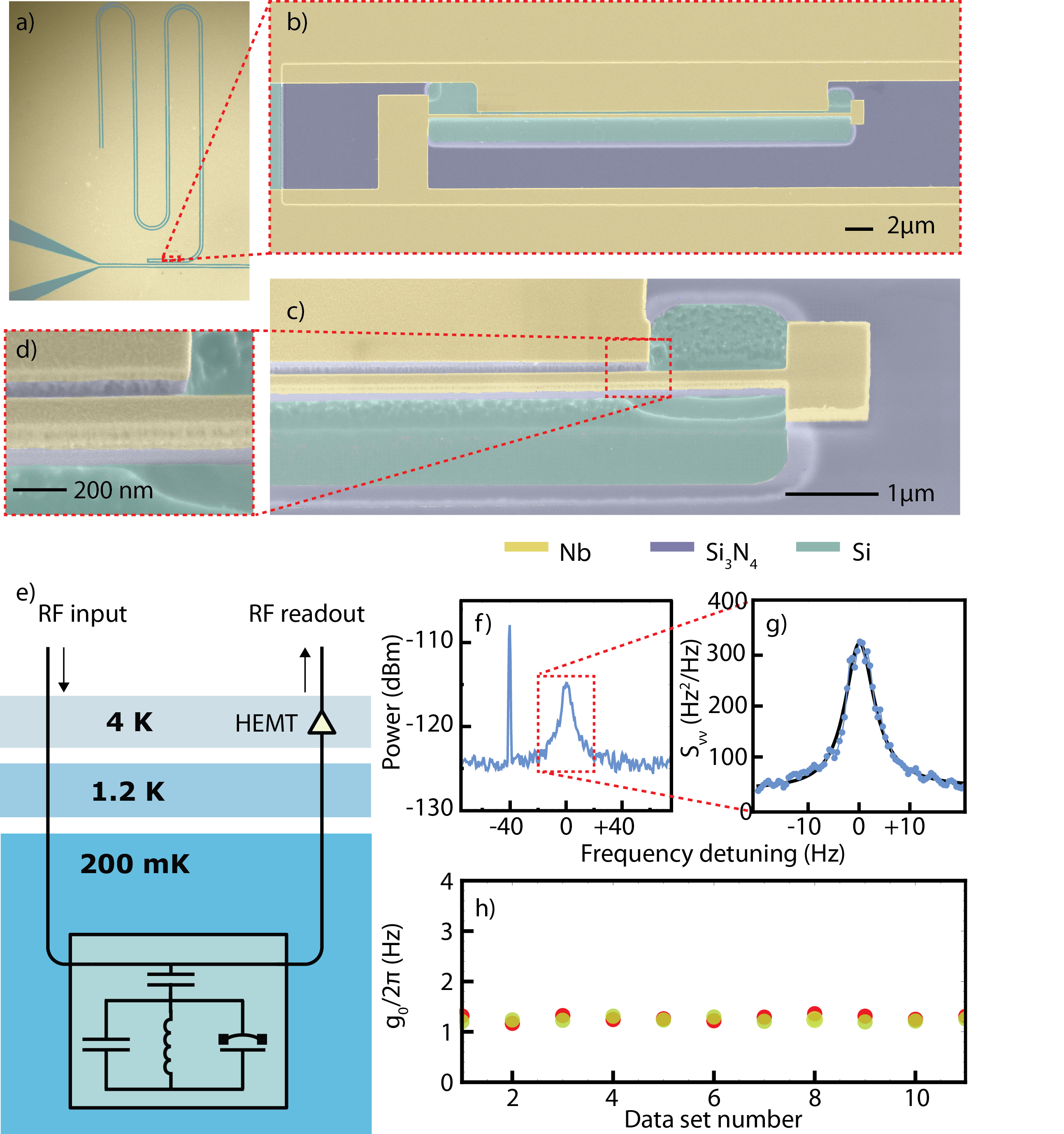}
\caption{Superconducting circuit nano-electromechanical system. a) Falsed-colored
scanning electron microscopy$\,$(SEM) of a quarter-wavelength coplanar
waveguide (CPW) resonator, coupled to a CPW feedline. The Nb center
stripline and ground are shown in light grey. The $6\ensuremath{\,\mu}$m
gaps are patterned by etching through the Nb layer$\,$(brown) down
to the Si$\,$(green). b) False-colored SEM$\,$(brown: Nb, green:
Si, violet: $\mathrm{Si_{3}N_{4}}$) of a $30\,\mathrm{\mu m}$ long
mechanical beam integrated in the microwave cavity. Note that the
beam investigated in this work is of dimensions $60\,\mathrm{\mu m}\,\times\,140\mathrm{\ensuremath{\,}nm\,}\times\,200\ensuremath{\,}\mathrm{nm}$.
c) and d) A zoomed-in view, showing the mechanical beam released from
the Si substrate. e) Measurement setup with a frequency modulated
pump tone on cavity resonance with $\Omega_{\textrm{mod}}=\Om-2\text{\ensuremath{\pi}}\times40$$\,$Hz
and a modulation index of $2\pi\times80\,\mathrm{{Hz}}/\Omega_{\mathrm{mod}}\approx5.5\times10^{-5}$as
the RF input, and a homodyne RF readout after signal amplification
at 4$\,$K with a high electron mobility transistor$\,$(HEMT). f)
Mechanical thermal noise spectrum and the calibration peak as measured
at the Q output of the mixer. g) Calibrated frequency noise spectral
density $S_{\nu\nu}$$\,$(blue dots, $\nu\equiv\frac{\Omega}{2\pi}$)
of the mechanical beam, and Lorentzian fit$\,$(black). h) The vacuum
coupling rate $g{}_{0}$ derived from two groups$\,$(red and yellow)
of measurement, with a 4$\,$dB power difference in the modulated
input tone. }

\label{f:sem} 
\end{figure}

We investigate these phenomena in a Nb superconducting circuit nano-electromechanical
system (similar in geometry to the ones studied in Refs. \cite{Regal2008,Rocheleau2010})
consisting of a quarter-wavelength coplanar waveguide$\,$(CPW) resonator
\cite{day_broadband_2003}$\,$(Figure\ \ref{f:sem}), parametrically
coupled to a nano-mechanical oscillator, consisting of a stoichometric,
high stress $\mathrm{Si_{3}N_{4}}$ beam coated with Nb. The microwave
resonator studied in this work exhibits a fundamental resonance frequency
of $\omega_{c}=2\pi\times6.07\,$GHz and has a linewidth $\kappa=2\pi\times742\,$kHz
of which $\kappa_{\mathrm{{ex}}}=\eta_{\mathrm{{c}}}\kappa=2\pi\times338\,\mathrm{{kHz}}$
are due to external coupling to the feedline. The Nb/$\mathrm{Si_{3}N_{4}}$
composite nano-mechanical beam has dimensions of $60\ensuremath{\,\mu\mathrm{m}}$$\times140\ensuremath{\,}$nm$\times$$200\ensuremath{\,}$nm
and shows at cryogenic temperatures very low dissipation$\mathrm{}$($Q_{\mathrm{m}}>10^{5}$),
with a damping rate of $\Gm=2\pi\times12\,$Hz resonating at $\Om=2\pi\times1.45\,$MHz.
This system thus resides in the resolved sideband regime as $\Om>\kappa$.
The thermal decoherence rate of the mechanical oscillator is $\Gm\bar{n}_{\mathrm{m}}=2\pi\times21$$\,$kHz.
Here, $\bar{n}_{\mathrm{m}}$ is the thermal equilibrium phonon occupancy
at the dilution refrigerator temperature of ca. 200 mK. This temperature
is far below the superconducting transition temperature of Nb$\,$(9.2$\,$K)
and significantly suppresses the thermal excitation of the microwave
cavity, as $\hbar\omega_{c}/\kB=288\,$mK, where $\hbar$ is the reduced
Planck constant and $\kB$ is the Boltzmann constant.

The interaction between the mechanical oscillator and the microwave
coplanar waveguide resonator is formally equivalent to the optomechanical
interaction \cite{Marquardt2009,Kippenberg2008} and quantified by
the vacuum coupling rate $g_{0}$ \cite{Gorodetsky2010} in the corresponding
interaction Hamiltonian 
\begin{equation}
\Hint=\hbar g_{0}(\hat{a}_{\mathrm{m}}+\hat{a}_{\mathrm{m}}^{\dagger})\hat{n}_{\mathrm{{p}}}
\end{equation}
 where $\hat{n}_{\mathrm{{p}}}$ is the intra-cavity photon number
operator, and $\hat{a}_{\mathrm{m}}$,  $\hat{a}_{\mathrm{m}}^{\dagger}$
are the ladder operators of the mechanical oscillator. The vacuum
coupling rate $g{}_{0}$ is the product of the electro-mechanical
frequency pulling parameter $G=\frac{\textrm{d}\omega_{c}}{\textrm{d}x}$,
denoting the cavity resonance frequency change upon mechanical displacement,
and the mechanical resonator's zero-point fluctuation $\xzp=\sqrt{\frac{\hbar}{2\meff\Om}}\approx30\,\mathrm{{fm}}$,
where $\meff\approx7\,\mathrm{{pg}}$ is the effective mass of the
beam. The coupling rate $g{}_{0}$ is calibrated by applying a known
frequency modulation to a microwave tone coupled into the cavity \cite{Gorodetsky2010}$\mathrm{}$(Figure\ \ref{f:sem}d)).
The spectrum of the homodyne readout is shown in Figure \ref{f:sem}e)
and f), yielding a measured coupling rate of $g{}_{0}=2\pi\times(1.26\pm0.05)\,\mathrm{{Hz}}$,
independent of the microwave power reaching the coplanar waveguide
cavity after several stages of attenuation in the refrigerator. %

\begin{figure}[tbph]
\centering \includegraphics[scale=0.42]{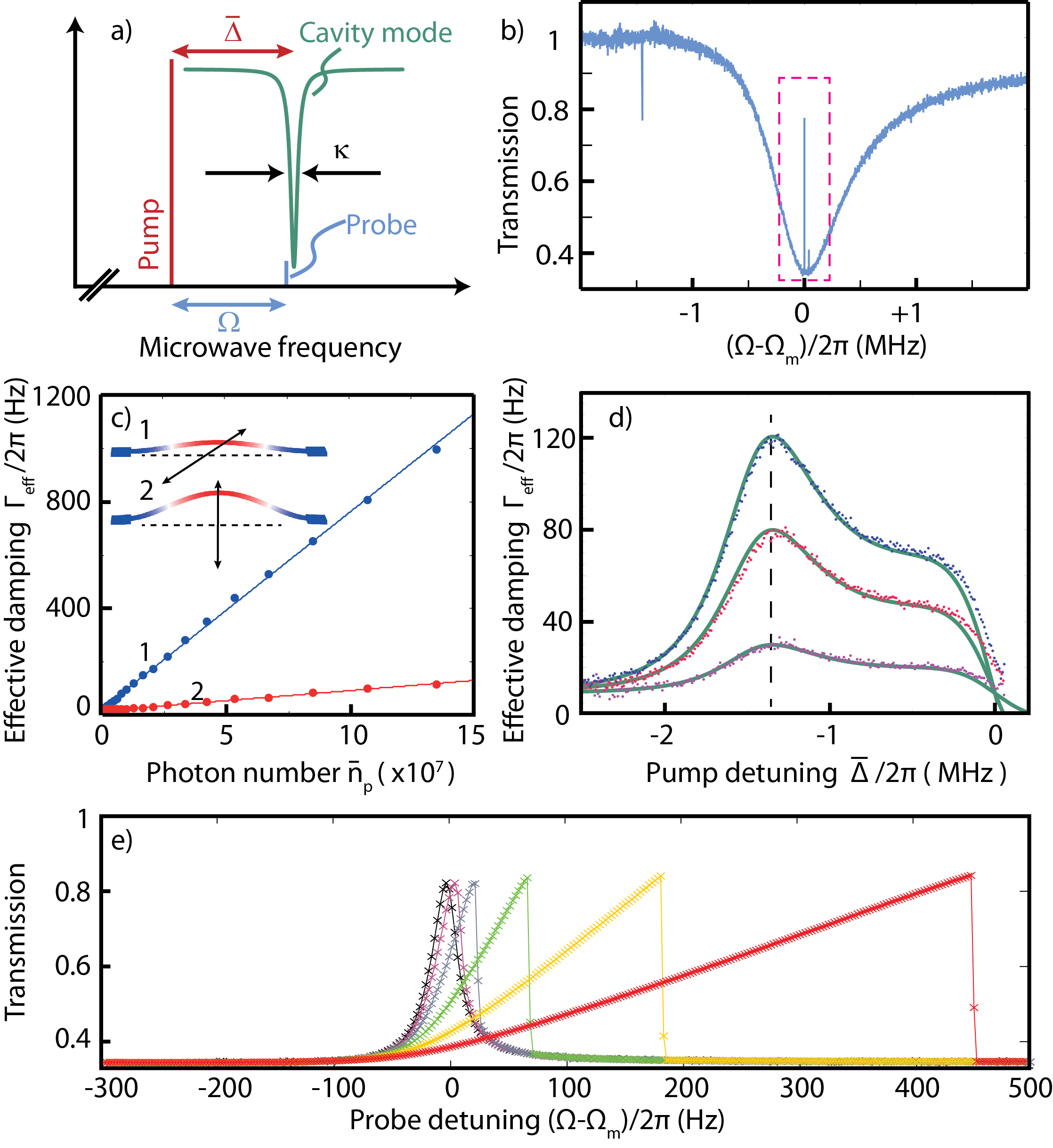}
\caption{Electromechanical and nano-mechanical response. a) The frequency of
the pump tone is detuned by $\bar{\Delta}$ from the cavity resonance
frequency. The probe tone has an offset frequency $\Omega$ from the
pump tone. The probe tone is tuned over the cavity resonance which
has a linewidth of $\kappa$. b) The probe transmission in the presence
a pump tone, as $\bar{\Delta}\approx\Om$. Two transparency windows
are observed, in the highlighted red regtangular, corresponding to
the in-plane and out-of-plane mechanical oscillation. c) Damping rate
versus intra-cavity pump photon number with pump detuning $\bar{\Delta}=-\Om$$\,$(blue:
in-plane mode, red: out-of-plane mode) together with linear fits$\mathrm{}$($\Gm\propto\np$
\cite{Kippenberg2005}). The probe tone has a power of -99$\,$dBm.
d) Damping rate versus pump detuning$\,$(points) and fits$\,$(lines)
according to the model of dynamical backaction. The pump tone has
constant powers of -57$\,$dBm$\,$(blue), -59$\,$dBm$\,$(red),
and -64$\,$dBm$\,$ (purple), which corresponds to intra-cavity photon
numbers of $1.1\times10^{7}$, $7.2\times10^{6}$, $2.3\times10^{6}$
at the maximum dynamical backaction damping, when the pump tone is
optimally detuned$\,$($\bar{\Delta}_{\mathrm{opt}}=-\sqrt{\Om^{2}+(\kappa/2)^{2}}$).
The probe tone has a constant power of -109$\,$dBm. e) Electromechanically
induced transparency with Duffing response. Intra-cavity pump photon
number is $3.9\times10^{7}$, powers of the probe tone are -99, -95,
-91, -87, -83, and -79$\,$dBm, respectively.}

\label{f:omit} 
\end{figure}

In our system, the effective radiation pressure force that is reflected
by the electromechanical interaction Hamiltonian$\,$($\hat{F}=i[\Hint,\hat{p}]/\hbar$),
gives rise to a modification of the dynamics of the mechanical oscillator.
Moreover, it leads to an interaction between two microwave fields
sent simultaneously into the cavity \cite{Teufel2011,massel_microwave_2011,Agarwal2010}.
This latter phenomenon which is also referred to as electromechanically
induced transparency arises as the overall radiation pressure of the
two fields, a strong pump at frequency $\omega_{\mathrm{p}}$ and
a weak probe field at frequency $\omega_{\mathrm{p}}+\Omega$, drives
the motion of the mechanical oscillator at the two fields' beat frequency
$\Omega$. The oscillation is resonantly enhanced if the frequency
difference $\Omega$ between the two fields coincides with $\Om$.
The driven motion, in turn, generates Stokes- and anti-Stokes sidebands
on the pump field, which can interfere with the probe field leading
to an induced transparency (or amplification for a blue detuned pump
\cite{massel_microwave_2011}). The resulting transmission coefficient
of the probe field is given by

\begin{equation}
t_{\mathrm{p}}=1-\frac{1}{2}\frac{1+if(\Omega)}{-i(\bar{\Delta}+\Omega)+\kappa/2+2\bar{\Delta}f(\Omega)}\eta_{c}\kappa,\label{eq:tp}
\end{equation}
 where $f(\Omega)=\hbar g_{0}^{2}\bar{a}^{2}\chi(\Omega)/\left(i(\bar{\Delta}-\Omega)+\kappa/2\right)$
and the susceptibility of the mechanical oscillator $\chi(\Omega)=\meff^{-1}(\Om^{2}-\Omega^{2}-i\Gm\Omega)^{-1}$.
Here, $\bar{a}^{2}=\np=\langle\hat{n}_{\mathrm{{p}}}\rangle$ denotes
the classical expectation value of the intra-cavity pump photon number
and $\bar{\Delta}=\omega_{\mathrm{p}}-\omega_{\mathrm{c}}+G\bar{x}$
is the effective detuning of the pump field with the static mechanical
displacement $\bar{x}$. Figure \ref{f:omit} shows the measured probe
transmission in the presence of a red-detuned pump $\bar{\Delta}=-\Om$,
resulting in an induced transmission window, which coincides with
the microwave cavity resonance. The width of this window in the weak
coupling regime is given by the effective mechanical damping rate
$\Geff\approx\Gm+\Gdba$, with $\Gdba=4g_{0}^{2}\bar{n}_{\mathrm{p}}/\kappa$
resulting from radiation-pressure induced dynamical backaction \cite{Teufel2008,Rocheleau2010,Teufel2011b,Kippenberg2005,Schliesser2006,Gigan2006,Arcizet2006a}.

We study this response experimentally using a vector network analyzer
to generate the probe tone, and analyze its direct transmission, in
this case without the homodyne interferometer. A standard microwave
generator provides the pump tone which is simultaneously coupled into
the cavity (cf. the SI for a more detailed description of the employed
measurement setup). A systematic investigation of the electromechanical
effective damping as a function of pump detuning and pump intra-cavity
photon number shows excellent agreement with theory$\,$(Figure \ref{f:omit}
a, b). Importantly, these measurements, together with the independently
determined $g_{0}$, can be used to provide an independent calibration
of the intra-cavity pump photon number $\np$, and therefore the microwave
attenuation in the refrigerator before entering the microwave cavity$\,$(62dB
attenuation in this measurement setup).

Interestingly, the nano-electromechanical system shows significant
deviations from the standard electromechanically induced transparency
behavior already for moderate pump power ($\np=3.9\times10^{7}$)
if the probe power is increased to more than $-91\,\mathrm{{dBm}}$$\,$(Figure
\ref{f:omit}c). Strongly asymmetric lineshapes of the transmission
window are observed. This asymmetry increases with an increasing probe
power - much in contrast to the fully linear theory. This nonlinearity
is a direct consequence of the fact that the transmission window is
modified by scattering of photons induced by the mechanical oscillator.
The latter exhibits a Duffing nonlinearity \cite{Nayfeh1979,Kozinsky2006}
with a critical amplitude of ca. 2$\,$nm and a Duffing parameter
$\beta=1.2\times10^{12}\,\mathrm{N/m^{3}}$, which is compatible with
earlier experiments \cite{Unterreithmeier2010}. This value is independently
determined in a set of measurements, in which the oscillator is driven
by an externally applied low-frequency AC voltage via the DC port
of a bias-tee$\,$(see SI). The full electromechanical dynamics$\,$(including
the Duffing nonlinearity for $\beta\neq$0) are captured by the set
of  equations in the Fourier domain : 

\begin{equation}
A\left(-i(\Omega+\bar{\Delta})+\frac{\kappa}{2}\right)=-i\bar{a}g_{0}X/\xzp+\sqrt{\eta_{c}\kappa/2}\, S\label{eq:Duffing1}
\end{equation}
\begin{equation}
\meff(\Om^{2}-\Omega^{2}-i\Gm\Omega)X+3\beta|X|^{2}X=-2\hbar A\bar{a}g_{0}/\xzp\label{eq:Duffing}
\end{equation}
for the amplitude of the intra-cavity probe field $A$ and and the
mechanical oscillation $X$. Here, the resolved-sideband regime is
assumed$\,$($\Om\gg\kappa$), as well as a large mechanical quality
factor $Q_{m}\gg\bar{a}/A$, so that that the dynamic$\,$(resonant)
response of the mechanical oscillator $X\propto\bar{a}A$ is much
larger than the static displacement $\bar{x}\propto\bar{a}^{2}$$\,$(cf.
SI). Furthermore, $|S|^{2}$ denotes the power of the probe field
sent towards the cavity. As seen in Figure \ref{f:omit}c, the Duffing
nonlinearity is thus mapped directly onto the transmission of the
probe field, yielding nonlinear behavior for probe powers as low as
$-91\,\mathrm{dBm}$ sent towards the cavity. The resulting bistability
of the microwave transmission could be used for non-volatile memory
applications \cite{mahboob_bit_2008}.

For lower probe powers, the radiation-pressure induced oscillation
amplitude of the mechanical mode remains well below the threshold
for non-linear oscillations. In this regime, the transmission of the
probe field is well described by equation$\,$(\ref{eq:tp}). Importantly,
the presence of the pump beam does not only induce a strong modification
of the transmission of the probe field, but also leads at the same
time to a fast variation of the phase $\phi=\mathrm{arg(}t_{\mathrm{p}})$
of the transmitted probe field across the transmission window. This
can lead to significant group delays \cite{Schliesser2010,Weis2010,Safavi-Naeini2011,Jiang2011a},
in analogy to that achieved with the electromagnetically induced transparency
in atomic \cite{Phillips2001,Liu2001} and in solid state media \cite{Longdell2005}.
The delay is given by 
\begin{equation}
\tau_{\mathrm{g}}=\frac{\partial\phi}{\partial\Omega}
\end{equation}
 for a microwave probe pulse whose center frequency falls into the
transmission window. In particular, in the resolved-sideband case
and for red-detuned pumping$\,$($\kappa<\Om=-\bar{\Delta}$), the
group delay is given by$\,$(see SI and \cite{Safavi-Naeini2011})
\begin{equation}
\tau_{\mathrm{g}}\approx\frac{2\eta_{c}C}{(1+C)(1+C-\eta_{c})}\Gm^{-1},\label{eq:tg}
\end{equation}
 where $C=\Omega_{\mathrm{c}}^{2}/(\kappa\Gamma_{\mathrm{m}})$ denotes
the electromechanical cooperativity parameter \cite{Groblacher2009a}
with the coupling rate $\Omega_{\mathrm{c}}=2g_{0}\bar{a}.$ The group
delay reaches its maximum value

\begin{equation}
\tau_{\mathrm{g}}^{\mathrm{max}}=2(1-\sqrt{1-\eta_{c}})^{2}/\eta_{c}\Gm
\end{equation}
 as the cooperativity approaches $C=\sqrt{1-\eta_{c}}$.

\begin{figure}[tb]
\centering \includegraphics[scale=0.4]{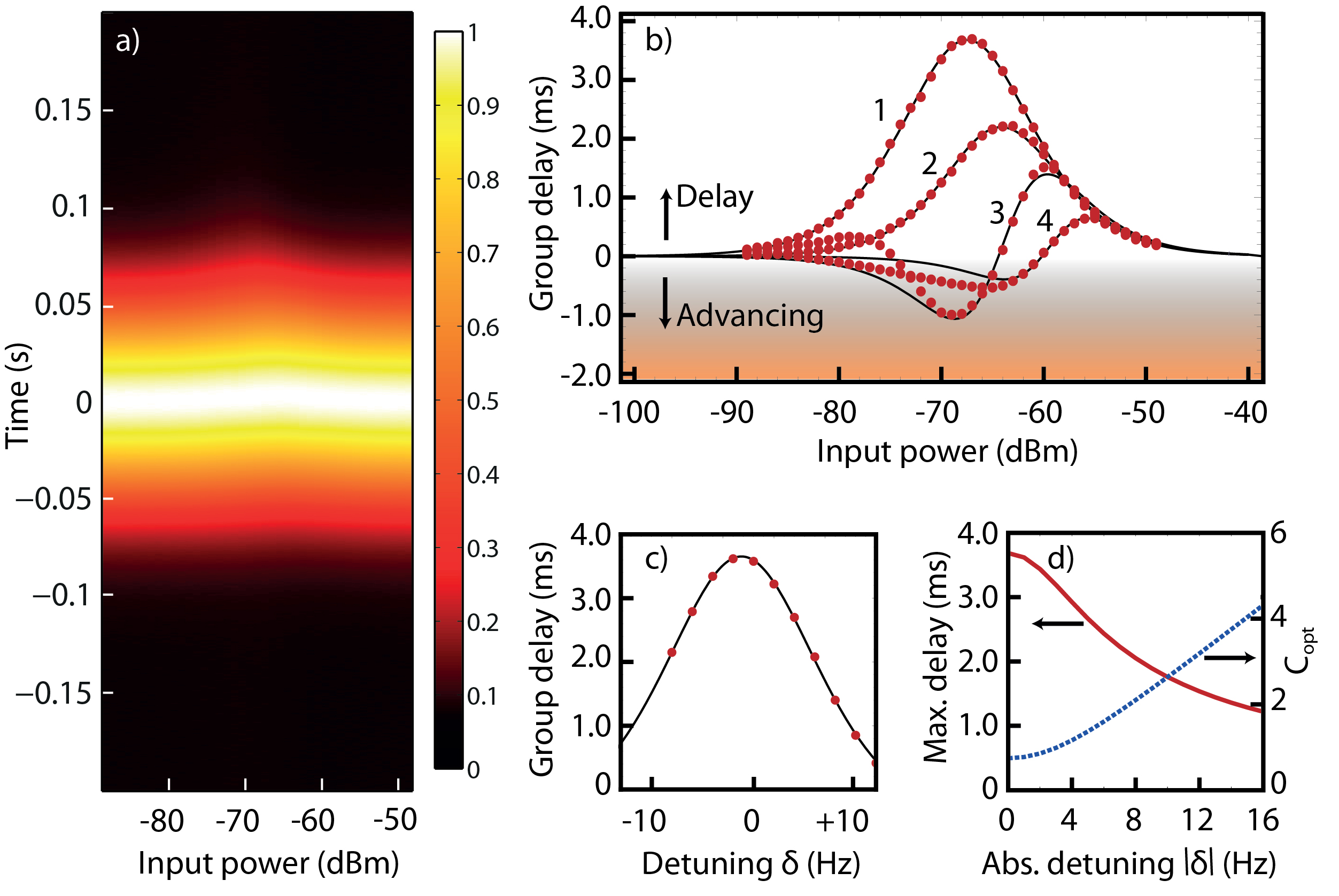} \caption{Sub- and superluminal microwave pulse propagation in a circuit electromechanical
system in the presence of electromechanically induced transparency.
a) Normalized transmittedamplitude of probe pulses, for various different
pump powers between -89 dBm to -49 dBm and tuned to the red sideband,
$-\bar{\Delta}=\Omega=\Om$. b) Extracted group delay for pump powers
between -89dBm to -49dBm, with various detunings $\vert\delta\vert=\vert\Omega-\Om\vert=2\pi\times(1.2\,\mathrm{{Hz}(1)},\,5.4\,\mathrm{{Hz}(2),}\,10\,\mathrm{{Hz}(3),}\,44\,\mathrm{{Hz}(4)})$.
The measured group delays - negative for superluminal and positive
for subluminal propagation - are plotted (red data points) versus
the input power together with the fittings from the full model$\,$(black).
c) Group delay for different detunings $\bar{\Delta}+\Om$ of the
pump, with $\Omega=\Om$ constant. The pump tone has a constant power
of -66 dBm. d) Maximum delay achievable for given frequency detuning
and optimum cooperativity. }

\label{f:tg} 
\end{figure}

To experimentally explore this predicted behavior, microwave probe
pulses are generated by modulating the amplitude of a weak$\,$(-108
dBm) probe tone derived from a microwave generator. The Gaussian-shaped
envelope functions$\,$(FWHM duration 83 ms) are generated with an
arbitrary waveform generator. The emission of a probe pulse is synchronized
with the acquisition of the transmitted probe field. Simultaneously,
a continuous-wave pump tone is sent to the cavity$\,$(see SI for
details). Figure \ref{f:tg} shows the results of these measurements.
A delay of the probe pulses can be observed when the power of the
pump is varied. The maximum group delay achieved is 3.5 ms with negligible
losses and pulse distortion as shown in Figure \ref{f:tg} a, b). 

In the case of a slightly detuned probe tone, $\delta=\Omega-\Om\neq0$,
a good approximation of the group delay is given by 
\begin{equation}
\tau_{\mathrm{g}}\approx-\frac{\frac{2\eta_{\mathrm{c}}}{1-\eta_{\mathrm{c}}}C\Gm}{4\delta^{2}+(1+C)^{2}\Gm^{2}}\cdot\frac{4\delta^{2}-\frac{(1-\eta_{\mathrm{c}}+C)(1+C)}{1-\eta_{\mathrm{c}}}\mathrm{\Gm^{2}}}{4\delta^{2}+\left(\frac{1-\eta_{\mathrm{c}}+C}{1-\eta_{\mathrm{c}}}\right)^{2}\Gm^{2}}
\end{equation}
This allows for a negative group delay, i.e. the advancing of the
microwave pulses, with sufficient detuning such that $\vert\delta\vert>\Gm/2$.
The probe delay measured when both pump tone power and probe tone
detuning are varied, reveals an excellent agreement with the full
theory (Figure \ref{f:tg} b, c, d).

\begin{figure}[t!]
\centering \includegraphics[scale=0.4]{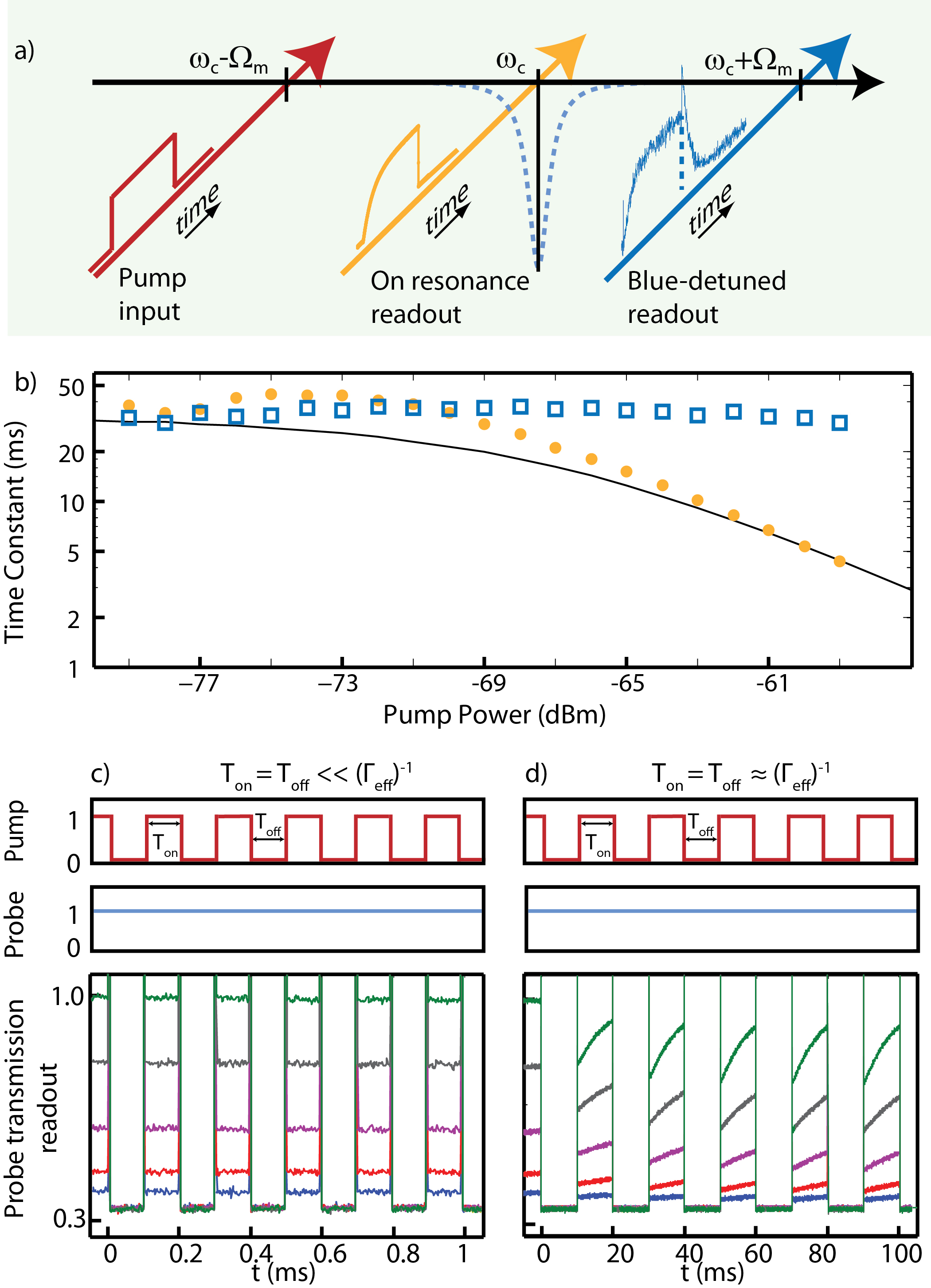} \caption{Switching dynamics. a) Measurement scheme, detail explained in the
text. b) Time constants of the ring-up$\,$(yellow dots) and ring-down$\,$(blue
squares) of the individual pulse as a function of pump power, with
the pulse ring-down measured blue-detuned$\,$($\bar{\Delta}=\Om$).
The curve$\,$(black) shows the calculation for the ring-up constant
with $\Gm=2\pi\times12\,$Hz. We expect here to have a smaller damping
rate than $\Gm=2\pi\times12\,$Hz with a lower measurement temperature
here around 100$\,$mK. c, d) A train of red-detuned$\,$($\bar{\Delta}=-\Om$)
pulses is sent into the cavity, with a period of $T_{\mathrm{on}}=T_{\mathrm{{off}}}=200$$\,$$\mathrm{\mu s}$$\,$(c)
and $T_{\mathrm{{on}}}=T_{\mathrm{{off}}}=20$$\,$ms$\,$(d), with
a power descending from -63 dBm$\,$(green) to -79 dBm$\,$(blue)
in steps of 4 dB. A weak probe tone$\,$(-99$\,$dBm) is present on
cavity resonance. Probe transmission on cavity resonance$\,$($\ocavity$)
measured with a spectrum analyzer in zero span mode is plotted versus
time, showing the dynamics of electromechanically induced transparency
on cavity resonance. }

\label{f:switching} 
\end{figure}

For a number of advanced optomechanical protocols for both quantum
and classical applications \cite{Verhagen2011,Romero-Isart2011,Wang2011a},
it is necessary to dynamically tune the coupling rate 
\begin{equation}
\Omega_{\mathrm{c}}(t)=2g_{0}\bar{a}(t).
\end{equation}
For switching applications, the response of the system can be limited
both by the dynamics of the mechanical mode amplitude$\,$($X(t)$)
as well as the pump field$\,$($\bar{a}(t)$). In the following, we
explore these dynamics experimentally. To this end, the pump tone
is tuned to the red sideband $\,$($\bar{\Delta}=-\Om$), and switched
by modulating its amplitude with rectangular pulses of $T_{\mathrm{{on}}}=150\,\mathrm{{ms}}$
duration. The probe tone, tuned in resonance with the cavity$\,$($\Omega+\omega_{\mathrm{p}}=\ocavity$),
is constantly on, and its transmission is recorded. While the intra-cavity
pump power rings up on a time scale $\kappa^{-1}\approx0.2\,\mathrm{\textrm{\ensuremath{\mu}}s}$,
the transmission of the probe beam builds up only at a significantly
smaller rate of $\Geff$, at which the mechanical oscillation amplitude
converges towards its steady-state value$\,$(see SI). Figure \ref{f:switching}b)
shows the characteristic timescales for probe transmission variations
as a function of pump power. Upon switching off of the pump field,
the probe transmission immediately drops, as the pump field$\,$(giving
rise to destructive interference) decays from the resonator at a fast
timescale of $\kappa^{-1}$. It is important to note that after switching
off the pump and the corresponding field enhanced optomechanical coupling
rate $\Omega_{\mathrm{c}}(t)=0$, the mechanical oscillation still
prevails, and consequently a small fraction of the probe field is
scattered to different$\,$(Stokes and anti-Stokes) frequencies. It
can be shown that the modification of the transmission$\,$(with respect
to the case where the coherent excitation of the mechanical oscillator's
amplitude has vanished) is negligible, since in this regime the change
in transmission due to the finite amplitude of the mechanical oscillator
is given by $\Delta|t_{\mathrm{p}}|^{2}/|t_{\mathrm{p}}|^{2}\approx\mathcal{O}(\varepsilon^{2})$,
with $\varepsilon=g_{0}x/\xzp\Om$ and $x$ being the amplitude of
motion. For the amplitudes concerned in this work, $\varepsilon$
is typically at the level of <1 \%. However, once the mechanical oscillation
is excited to its steady state amplitude by the beating of the pump
and probe fields, it decays only at a very slow rate of $\Gm$ in
the absence of the pump beam. Therefore, advantageously for the switching,
if a series of pump pulses is applied, whose period $T=T_{\mathrm{{on}}}+T_{\mathrm{{off}}}$
is much shorter than this timescale, the transmission of the probe
follows the pump modulation, determined by the decay time $\kappa^{-1}$
of the microwave cavity (Figure \ref{f:switching}e). We show the
counterintuive regime where the switching time can be substantially
faster than the slowest timescale in the problem (i.e. the inverse
effective mechanical damping rate, $\Geff^{-1}$) and is limited only
by the microwave cavities decay time ($\kappa^{-1}\ll\Geff^{-1}$
). In a regime of intermediate pulse periods, the mechanical amplitude
partially decays between the pump pulses, leading to a slower recovery
of the full probe transmission when the pump is switched back on.
Switching is possible if the mechanical mode does not decay significantly
between the individual pump pulses$\,$($\kappa^{-1}<T_{\mathrm{{off}}}\lesssim\Gm^{-1}$).
During longer off-times$\,$($T_{\mathrm{{off}}}>\Gm^{-1}$), the
mechanical oscillator relaxes towards its equilibrium position, however
$T_{\mathrm{{on}}}>\Geff^{-1}$ suffices already to drive it back
to its steady state oscillation, at which the probe tone has a transmission
transparency.

Note, that while in the above electromechanical-switching experiments
the pump tone is modulated, and the probe tone remains stationary,
utilizing the demonstrated temporal control in conjuction with probe
pulses enables to achieve storage of pump pulses in the mechanical
oscillator via conversion of the pump pulse into a coherent excitation
of the mechanical oscillator \cite{Braunstein2003,Fiore2011}. 

In conclusion, we have demonstrated that electromechanically induced
transparency can be used in electromechanical systems to manipulate
the transmission and delay of a microwave signal in a fully integrated
architecture without the need of photon detection and regeneration.
Interestingly, the switching can be faster than the timescale of the
mechanical oscillator's energy decay. Using electromechanically induced
transparency, both classical microwave signals can be switched, or
routed \cite{hoi_demonstration_2011} through arrays of electromechanical
systems, while the associated delays enable synchronization of microwave
pulses. Both delay and advancing of pulses have been achieved. The
employed microfabrication techniques offer in this context a far-reaching
flexibility in the design of both mechanical and microwave properties--including
carrier frequencies and bandwidth--as well as the overall architecture
of complex networks. The delay and advancing of pulses may also be
extended to single microwave photon pulses \cite{Houck2007}. The
prerequisites for preserving the single photon state is that the thermal
decoherence time ($1/\Gamma_{m}\bar{n}_{m}$) is long compared to
the photon delay/advance time. The pulse duration (i.e. bandwith of
the pulses) is limited by the width of the transparency window ($\Gamma_{\mathrm{eff}}$)
to avoid pulse distortion, implying that the condition $\Gamma_{\mathrm{eff}}\gg\bar{n}_{\mathrm{m}}\Gm$
needs to be satisfied. In the weak coupling limit ($\Omega_{\mathrm{c}}<\kappa$),
the latter is equivalent to a cooperativity $C$ exceeding the thermal
occupation $\bar{n}_{\mathrm{m}}$.

While our system already reaches coupling rates exceeding the mechanical
decoherence rate $\Omega_{\mathrm{c}}\gtrsim\bar{n}_{\mathrm{{m}}}\Gm$
even at a moderate cryogenic temperature, a larger $g_{0}$ \cite{Teufel2011,Verhagen2011},
or a simple improvement in the measurement setup, allowing a larger
pump field (sustainable by the employed Nb cavities) would in principle
already be sufficient for the system to reside in the coherent coupling
regime $\Omega_{\mathrm{c}}>(\bar{n}_{\mathrm{{m}}}\Gm,\kappa)$.
Combining the system with the powerful advances in the generation
and detection of single microwave photons \cite{Houck2007}, thereby
may enable control over the propagation of nonclassical states using
the electromechanical architecture, and allow for complete storage
and retrieval of a microwave quantum state in long-lived mechanical
excitations.

\subsection*{Methods:}

1. Nano-electromechanically coupled system: For the coplanar waveguide
(CPW) structures a characteristic impedance of $Z_{0}\approx50\,\Omega$
is realized with a $10\,\mu$m wide center stripline separated by
a gap of $6\ensuremath{\,\mu}$m from the ground plane$\,$(see Figure$\,$\ref{f:sem}).
The one-sided cavity with a typical length of $5.3\ensuremath{\,}$mm
is capacitively coupled to a CPW feedline on one end, and shorted
on the other end. This CPW cavity is patterned into a Nb thin film
deposited on top of a Si substrate. Frequency multiplexing is realized
by embedding on a single chip several cavities of different length
coupled to a single CPW feedline. The nano-mechanical object integrated
in the cavity is a high-aspect-ratio beam, $60\ensuremath{\,\mu}$m
long, $140\ensuremath{\,}$nm wide and has a thickness of $200\ensuremath{\,}$nm,
which consists of $130\ensuremath{\,}$nm thick Nb on top of $70\,$nm
thick tensile-stressed $\mathrm{Si_{3}N_{4}}$. The high tensile-stress
of $\mathrm{Si_{3}N_{4}}$ overcomes the compressive stress of Nb.

2. Low temperature measurement setup: All experiments have been performed
in a dilution refrigerator at around 200 mK. In the dilution refrigerator
there are three signal lines: two coaxial-cable lines connecting the
two ends of the sample feedline for the microwave input- and output-signal,
and a low-frequency line to carry a drive signal $\,$($<5\,$MHz)
that enables resonant excitation of the nano-mechanical beam via Coulomb
force. A sample of dimension 10$\,$mm$\times$ 6$\,$mm is mounted
in a gold-plated copper box. SMA coaxial connector at each end is
silver-glued to the CPW feedline, transmits signals to/from the sample.
Thermal noise from the tones at room temperature is suppressed using
attenuators at successive temperature stages, and by the inertial
attenuation of the coaxial cable. The DC-block filter besides the
sample prevents the DC current in the overall transmission loop. A
HEMT amplifier is anchored at 4K and is isolated from the sample output
by a circulator. For more details of the individual measurements,
please refer to the SI.

\subsection*{Acknowledgements:}

TJK acknowledges support support by the NCCR of Quantum Engineering
and an ERC Starting Grant (SiMP) and the Swiss National Science Foundation
(SNF). Financial support from the German Excellence Initiative via
the \textquoteleft{}\textquoteleft{}Nanosystems Initiative Munich\textquoteright{}\textquoteright{}
(NIM) is gratefully acknowledged. Samples were grown and fabricated
at the Center of MicroNanotechnology (CMi) at EPFL. The authors acknowledge
the assistance of Stefan Weis, Thomas Niemczyk and Haytham Chibani
in fabrication, Pertti Hakonen and Pasi L\"{a}hteenm\"{a}ki for measurement
in the early phase of the project. 

\subsection*{Author contributions:}
XZ designed and fabricated the samples. The cryogenic measurement setup was implemented
by FH and HH. FH, XZ, HH and AS performed the experiments. XZ and AS performed theoretical
modeling and analysis of the data. XZ wrote the paper with
guidance from AS and TJK.  All authors discussed the results and contributed to the final version of the manuscript.

\clearpage

\renewcommand{\thefigure}{S\arabic{figure}}
\renewcommand{\thetable}{\textbf{S}\arabic{table}}
\renewcommand{\theequation}{$\mathrm{S\,} $\arabic{equation}}
\setcounter {figure} {0}
\setcounter {equation} {0}
\setcounter {table} {0}
\begin{widetext}
\part{Supplementary Information -Control of microwave signals using circuit nano-electromechanics}

\section{determination of the vacuum coupling rate }

The vacuum optomechanical coupling rate $g_{0}$ quantifies the strength
of electromechanical coupling, analogous to cavity quantum electrodynamics.
Assuming $\nm=\kB T/\hbar\Om\gg1$$\,$(with $\kB$: Boltzmann constant,
$T$: temperature, $\hbar$: reduced Planck constant, $\Om$: mechanical
frequency), the spectral density of fluctuations of the cavity resonance
frequency induced by mechanical displacement are given by \cite{Gorodetsky2010}
\begin{equation}
S_{\omega\omega}(\Omega)=G^{2}S_{xx}(\Omega)\approx g_{0}^{2}\frac{2\Om}{\hbar}\frac{2\Gm\kB T}{(\Omega^{2}-\Om^{2})^{2}+\Gm^{2}\Omega^{2}}\label{eq:Sww}
\end{equation}
where $G$ is the frequency pulling parameter, and $\Gm$ is the mechanical
damping rate. The integration of equation$\,($\ref{eq:Sww}) gives 

\begin{equation}
\langle\delta\ocavity^{2}\rangle=\intop_{-\infty}^{+\infty}S_{\omega\omega}(\Omega)\frac{d\Omega}{2\pi}=S_{\omega\omega}(\Om)\frac{\Gm\pi}{2\pi}=2\langle\nm\rangle g_{0}^{2}\label{eq:g0}
\end{equation}
where $\bar{n}_{\mathrm{m}}$ is the mean mechanical occupancy. The
fluctuations of the detected homodyne signal can be expressed as 
\begin{equation}
S_{II}^{\mathrm{meas}}(\Omega)=\frac{2K(\Omega)}{\Omega^{2}}S_{\omega\omega}(\Omega)\label{eq:SII}
\end{equation}
where the transduction function $K(\Omega)\equiv K(\Omega,\bar{\Delta},\eta_{c},\kappa,\np)$
is calibrated by applying a frequency modulation of the pump tone
at frequency $\Omega_{\textrm{mod}}=\Om-2\text{\ensuremath{\pi}}\times40$$\,$Hz,
with a modulation index of $\phi_{0}=2\pi\times80\mathrm{}\mathrm{Hz}/\Omod\approx5.5\times10^{-5}$.
As the effective noise bandwidth chosen in the spectral analyzer is
much narrower than the spectral features of the mechanical mode, and
assuming the signal at $\Omod$ is dominated by the modulation, the
transduction function $K(\Omega)$ can then be calibrated in absolute
terms at this frequency, via the relation

\begin{equation}
K(\Omod)\approx2S_{II}^{\mathrm{meas}}(\pm\Omod)\textrm{\ensuremath{\frac{\textrm{ENBW}}{\phi_{0}^{2}}}}\label{eq:Kmod}
\end{equation}
where ENBW is the bandwidth given in direct frequency, i.e. in Hz.
With the homodyne detection at zero detuning$\,$($\Delta=0$), i.e.
the pump tone $\opump$ is on cavity resonance $\ocavity$, in this
experiment, the transduction function $K(\Omega)$ is sufficiently
constant$\,$($K(\Omega)=K(\Omega_{\mathrm{mod}})\approx K(\Omega_{\mathrm{m}})$)
over the relevant range of frequencies. With equation$\,$(\ref{eq:g0})-(\ref{eq:Kmod}),
the vacuum coupling rate is derived from the measurement, knowing
the mechanical occupancy number $\bar{n}_{\mathrm{m}}$, via \cite{Gorodetsky2010}

\begin{equation}
g_{0}^{2}=\frac{1}{2\langle\nm\rangle}\frac{\phi_{0}^{2}\Om^{2}}{2}\frac{S_{II}^{\mathrm{meas}}(\Om)}{S_{II}^{\mathrm{meas}}(\pm\Omod)}\frac{\Gm/4}{\textrm{ENBW}}
\end{equation}

\section{Sample Fabrication}

The fabrication starts with a Si substrate$\,$(Figure \ref{f:FabricationFlow}$\,$a)).
A high tensile-stressed layer of 100$\,$nm $\textrm{Si\ensuremath{_{3}}}\textrm{N}_{\textrm{4}}$
is deposited with low pressure chemical vapor deposition$\,$(LPCVD)
on top of the Si substrate$\,$(Figure. \ref{f:FabricationFlow}$\,$b)).
$\textrm{Si\ensuremath{_{3}}}\textrm{N}_{\textrm{4}}$ is patterned
into small patches using electron beam$\,$(e-beam) lithography, reactive
ion etching$\,$(RIE) and buffered hydrogen fluoride$\,$(HF) etching.
The mechanical beams will later sit on those patches$\,$(Figure.
\ref{f:FabricationFlow}$\,$c,d,e)). A thin Nb film of 130$\,$nm
is deposited on top of the substrate, and microwave cavities are patterned
with e-beam lithography and the Nb film is etched through with RIE$\,$(Figure.
\ref{f:FabricationFlow}$\,$f,g,h,i)). A short patch of Nb is left
on top of the $\textrm{Si\ensuremath{_{3}}}\textrm{N}_{\textrm{4}}$
patch, which is between the center and ground of the MW cavity CPWs.
At this exact location, the mechanical beam is patterned with e-beam
lithography and released in the following fabrication steps, i.e.,
anisotropic- and isotropic- RIE. $\,$(Figure. \ref{f:FabricationFlow}$\,$j,k,l))

\begin{figure}[htbp]
\centering \includegraphics[width=0.7\linewidth]{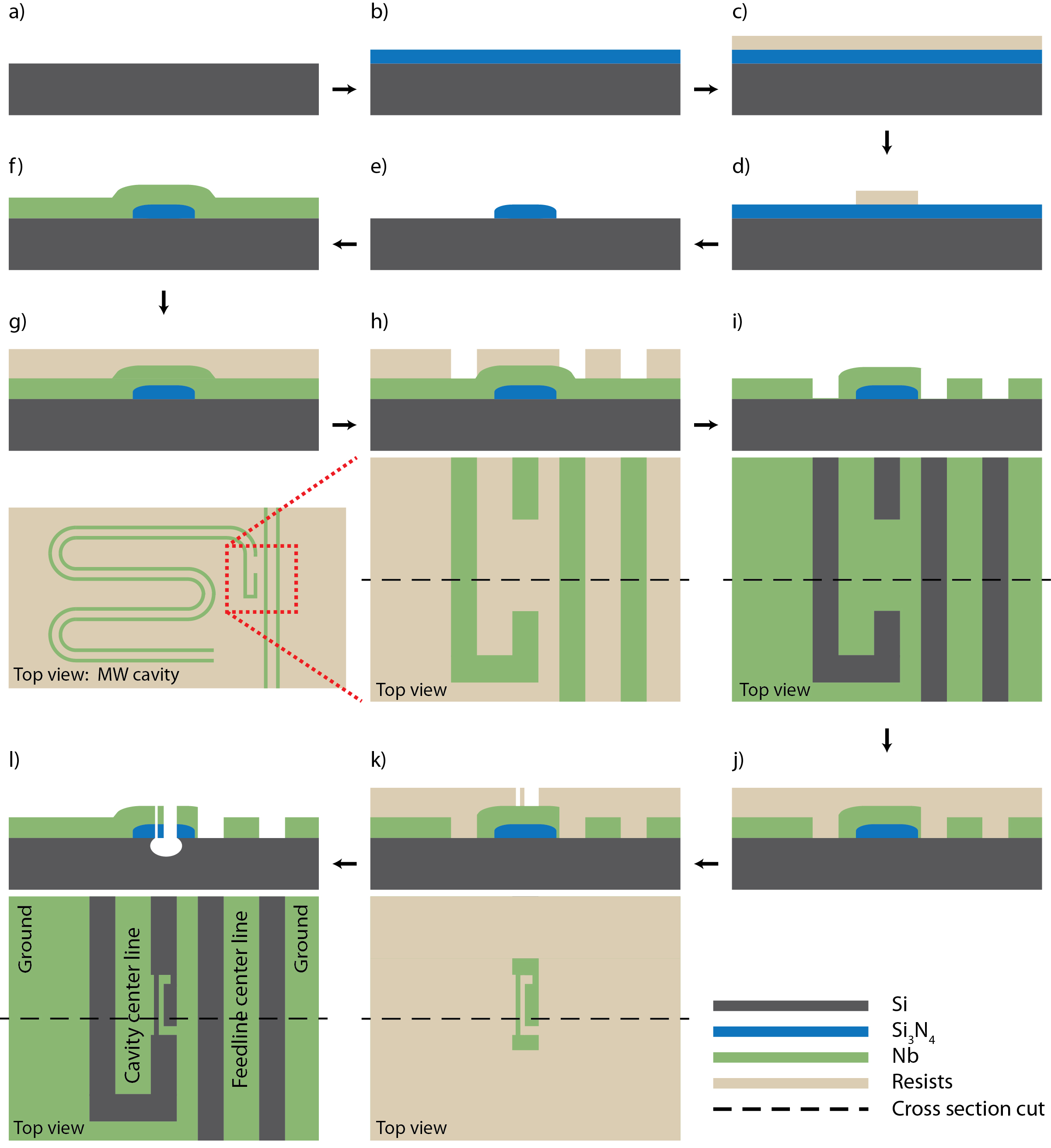}
\caption{Fabrication flow of the electromechanical system. a)$\,$Si substrate;
b)$\,$LPCVD deposition of 100nm $\textrm{Si\ensuremath{_{3}}}\textrm{N}_{\textrm{4}}$;
c)$\,$Negative resist coating; d)$\,$ E-beam patterning; e)$\,$RIE
and buffered HF etching of $\textrm{Si\ensuremath{_{3}}}\textrm{N}_{\textrm{4}}$;
f)$\,$Sputtering of 130nm Nb; g)$\,$Positive resist coating; h)$\,$E-beam
pattering of MW cavities; i)$\,$RIE of Nb; j)$\,$Positive resist
coating; k)$\,$E-beam pattering of mechanical beams; j)$\,$RIE of
Nb, $\textrm{Si\ensuremath{_{3}}}\textrm{N}_{\textrm{4}}$ and Si.}

\label{f:FabricationFlow} 
\end{figure}

\section{Measurement setups}

The experiments are carried out at cryogenic temperature around 200$\,$mK
in a dilution refridgerator. In the dilution refrigerator there are
three signal lines: two coaxial-cable lines connecting the two ends
of the sample feedline for the microwave input- and output-signal,
and a low-frequency line to carry a drive signal$\,$($<5\,$MHz)
that enables resonant excitation of the nano-mechanical beam via Coulomb
force. A sample of dimension 10$\,$mm$\times$6$\,$mm is mounted
in a gold-plated copper box. SMA coaxial connectors are silver-glued
to the coplanar wave guide$\,$(CPW) feedline at two ends, transmiting
signals to/from the sample. Thermal noise from the tones at room temperature
is suppressed using various attenuators at successive temperature
stages, and by the attenuation of the coaxial cable. The DC-block
filter$\,$(Minicircuits BLK-18 S+)) besides the sample prevents DC
currents in the overall transmission loop. A high electron mobility
transistor$\mathrm{}$(HEMT) amplifier is anchored to 4$\,$K and
is isolated from the sample output by a circulator$\,$(Pamtech CTH
1392KS)). 

For the OMIT measurement shown in Figure \ref{f:OMIT_MeasurementSetup}
a), a pump tone from a signal generator$\,$(R\&S SMF 100A) and a
probe tone from a network analyzer$\,$(R\&S ZVA8) are coupled at
room temperature and are transmitted down to the sample. 

To experimentally explore the delay and advance of microwaves, as
shown in Figure \ref{f:OMIT_MeasurementSetup} b), a continuous-wave
pump tone tuned to the red sideband$\,$($\Delta=-\Om$) is sent from
a signal generator$\,$(R\&S SMF 100A) to the cavity. A weak probe
tone tuned in resonance$\,$($\Omega=\Om$) with the cavity is generated
from a second microwave source$\,$(R\&S SMF 100A). The probe tone
is amplitude-modulated by a Gaussian-shaped envelope generated with
an arbitrary waveform generator$\,$(LeCroy arbstudio 1104D). The
emission of this pulse triggers the acquisition of the transmitted
probe field via an electronic spectrum analyzer$\,$(R\&S FSV-30)
in zero-span mode. The actual delay of the output pulse due to the
OMIT is compared to the pulse output without a pump tone present.

To study the dynamics of OMIT, the pump tone generated from a signal
generator$\mathrm{}$(R\&S SMF 100A) is tuned to the red sideband$\,$($\Delta=-\Om$),
and switched by modulating its amplitude with rectangular pulses generated
from an arbitrary wave generator$\,$(LeCroy arbstudio 1104D). The
probe tone generated from an independent microwave source$\,$(R\&S
SMF 100A), tuned in resonance with the cavity ($\Omega=\Om$), is
constantly on. An electronic spectrum analyzer in zero-span mode$\,$(R\&S
FSV-30) is triggered on with the first rectangular pulse, and the
probe transmission is recorded.

\begin{figure}[htbp]
\centering \includegraphics[width=1\linewidth]{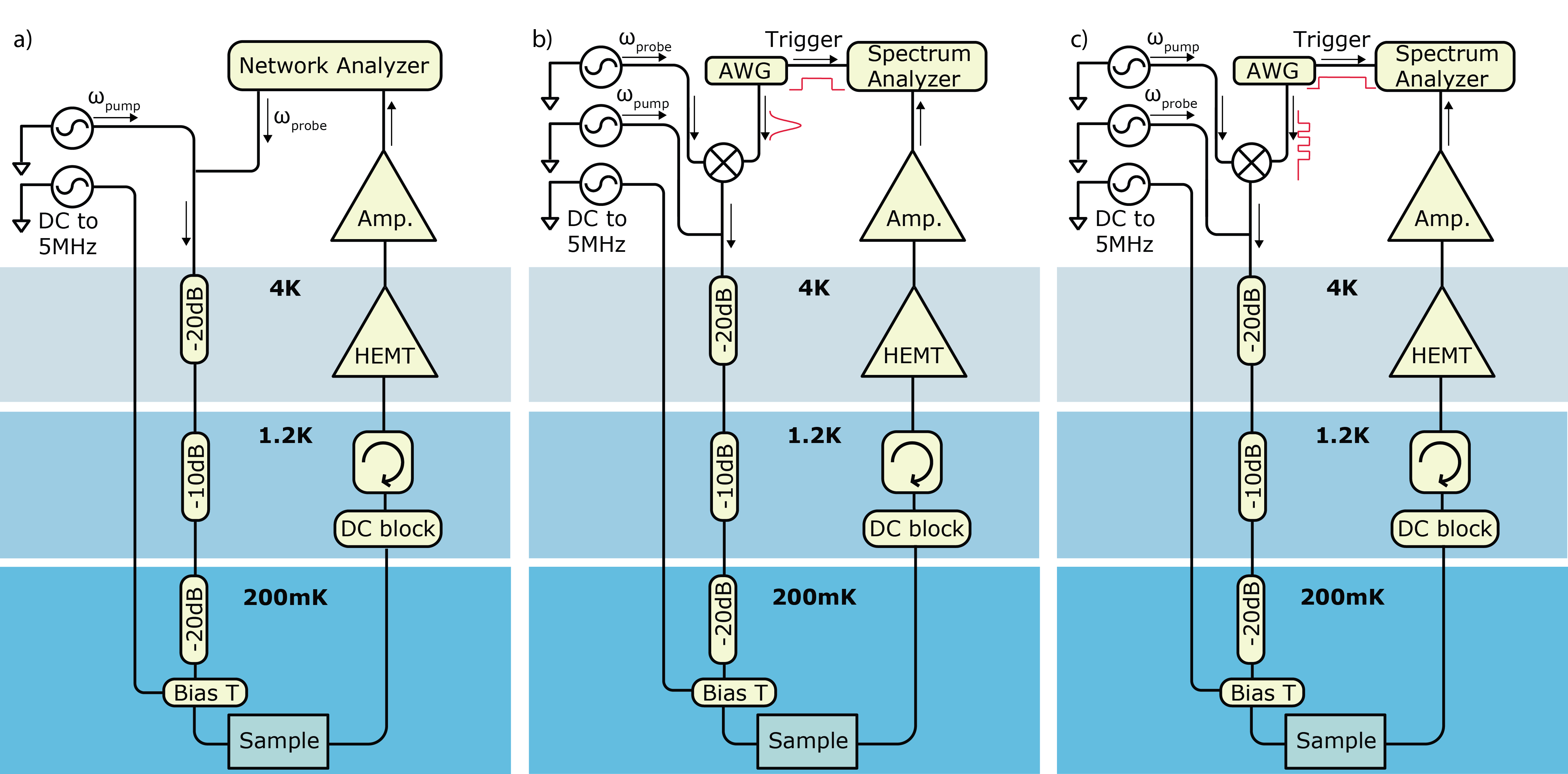}\caption{Measurement setups for a)$\,$OMIT, b)$\,$pulse delay measurement,
and c)$\,$pulse train dynamics }

\label{f:OMIT_MeasurementSetup} 
\end{figure}

\section{Model for mechanical Duffing nonlinearity}

In a frame rotation at the pump frequency, with $\Delta=\opump-\ocavity$,
neglecting quantum noise and thermal noise terms, we have

\begin{spacing}{0.5}
\begin{equation}
\frac{d}{dt}\ahat(t)=\left(i\Delta-\frac{\kappa}{2}\right)\ahat(t)-iG\xhat(t)\ahat(t)+\sqrt{\eta_{c}\kappa/2}s_{\mathrm{in}}(t)
\end{equation}
\begin{equation}
\frac{d}{dt}\xhat(t)=\frac{\phat(t)}{\meff}
\end{equation}

\end{spacing}

\begin{equation}
\frac{d}{dt}\phat(t)=-\meff\Om^{2}\xhat(t)-\hbar G\ahat^{\dagger}(t)\ahat(t)-\Gm\phat(t)-\beta\xhat^{3}(t)
\end{equation}
where $\ahat$ is the annihilation operators of the cavity mode, $\hat{x}$
and $\hat{p}$ are the position and momentum operators of the mechanical
degree of freedom having effective mass $\meff$ and Duffing nonlinearity
parameter $\beta$, and $s_{\mathrm{in}}(t)$ is the drive amplitude
normalized such that $\vert s_{\mathrm{in}}(t)\vert^{2}$ is the photon
flux at the input of the cavity. In the steady state, all time derivative
terms are zero, and the self-consistent static solutions for intra-cavity
field and mechanical displacement are:

\begin{spacing}{0.5}
\begin{equation}
\bar{a}=\frac{\sqrt{\eta_{\mathrm{c}}\kappa/2}}{-i(\Delta-G\bar{x})+\frac{\kappa}{2}}\bar{s}_{\mathrm{in}}
\end{equation}
\begin{equation}
\meff\Om^{2}\bar{x}+\beta\bar{x}^{3}+\hbar G\bar{a}^{2}=0
\end{equation}
In order to solve the equations in the presence of the pump and probe
fields $s_{\mathrm{in}}(t)=\bar{s}_{\mathrm{in}}+\delta s_{\mathrm{in}}(t)$
, we use the ansatz $\ahat(t)=\bar{a}+\delta\ahat(t)$, $\xhat(t)=\bar{x}+\delta\xhat(t)$,
and chose the phase reference of the cavity field so that $\bar{a}$
is real and positive. Then we have

\begin{equation}
\frac{d}{dt}\delta\ahat(t)=\left(i\Delta-iG\bar{x}-\frac{\kappa}{2}\right)\delta\ahat-iG\left(\bar{a}+\ahat(t)\right)\delta\xhat(t)+\sqrt{\eta_{\mathrm{c}}\kappa/2}\delta\mathrm{s}_{i\mathrm{n}}(t)\label{eq:eom1}
\end{equation}

\begin{equation}
\meff\left(\frac{d^{2}}{dt^{2}}\delta\xhat(t)+\Gm\frac{d}{dt}\delta\xhat(t)+\Om^{2}\delta\xhat(t)\right)=-\hbar G\bar{a}\left(\delta\ahat(t)+\delta\ahat^{\dagger}(t)\right)-\beta\left(\bar{x}^{3}+\delta\xhat^{3}(t)+3\bar{x}^{2}\delta\xhat(t)+3\bar{x}\delta\xhat^{2}(t)\right)\label{eq:eom2}
\end{equation}
For a given $\Omega=\oprobe-\opump$, and a probe field of $\delta s_{\mathrm{in}}(t)=s_{\mathrm{P}}e^{-i\Omega t}$,
we furthermore use the ansatz for the sideband fields of the pump
(one of which is the probe):

\begin{equation}
\delta a(t)=A^{-}e^{-i\Omega t}+A^{+}e^{+i\Omega t}
\end{equation}
\begin{equation}
\delta a^{*}(t)=(A^{+})^{*}e^{-i\Omega t}+(A^{-})^{*}e^{+i\Omega t}
\end{equation}
\begin{equation}
\delta x(t)=Xe^{-i\Omega t}+X^{*}e^{+i\Omega t}
\end{equation}
Sorting the rotation terms, the relevant equations with $e^{-i\Omega t}$
are given by

\begin{equation}
A^{-}\left(-i(\Omega+\Delta+G\bar{x})+\frac{\kappa}{2}\right)=-iG\bar{a}X+\sqrt{\eta_{\mathrm{c}}\kappa/2}s_{\mathrm{P}}
\end{equation}

\begin{equation}
A^{+}\left(i(\Omega-\Delta-G\bar{x})+\frac{\kappa}{2}\right)=-iG\bar{a}X^{*}
\end{equation}

\begin{equation}
\meff\left(\Om^{2}-\Omega^{2}-i\Gm\Omega\right)X=-\hbar G\bar{a}\left(A^{-}+(A^{+})^{*}\right)-3\beta(\bar{x}^{2}X+\vert X\vert^{2}X)
\end{equation}
In the resolved side band regime, the lower sideband $A^{+}$ is far
off-resonance and is much smaller than $A^{-}$, we therefore write
$A\approx A^{-}$and approximate $A^{+}\approx0$. Here, the resolved-sideband
regime is assumed$\,$($\Om\gg\kappa$), as well as a large mechanical
quality factor $\Om/\Gm\gg\bar{a}/A$, so that the dynamic$\,$(resonant)
response of the mechanical oscillator $X\propto\bar{a}A$ is much
larger than the static displacement $\bar{x}\propto\bar{a}^{2}$.
Taking into account that $\bar{x}\ll\delta\xhat(t)$, and using $\bar{\Delta}=\Delta+G\bar{x}$
where $G=g_{0}/\sqrt{\frac{\hbar}{2\meff\Om}}$, we have the pair
of equations in the main article:

\begin{equation}
A\left(-i(\Omega+\bar{\Delta})+\frac{\kappa}{2}\right)=-i\bar{a}g_{0}\sqrt{\frac{2\meff\Om}{\hbar}}X+\sqrt{\eta_{\mathrm{c}}\kappa/2}s_{\mathrm{P}}
\end{equation}
\begin{equation}
\meff(\Om^{2}-\Omega^{2}-i\Gm\Omega)X+3\beta\vert X\vert^{2}X=-2A\bar{a}g_{0}\sqrt{2\Om\hbar\meff}
\end{equation}
The transmission ratio of the probe field is given by:
\end{spacing}

\begin{equation}
t_{\mathrm{p}}=\frac{s_{\mathrm{P}}-\sqrt{\eta_{\mathrm{c}}\kappa/2}A}{s_{\mathrm{P}}}
\end{equation}

An independent calibration measurement of the Duffing parameter $\beta$
is carried out by a direct mechanical driving through the low-frequency
line coupled to the cavity via a bias-T. A vector networkanalyzer
centered around the mechanical frequency$\,$$\Om$ probes over a
200$\,$Hz wide spectrum. A probe tone of $\omega=\ocavity$ is sent
simultaneously into the cavity with a frequency modulation at $\Omod=\Om-2\text{\ensuremath{\pi}}\times80$$\,$Hz
and a modulation index of $2\pi\times100\,\mathrm{{kHz}}/\Omod$.
The frequency modulation from the driven mechanical oscillation is
calibrated with this injected frequency modulation, from the homodyne
readout spectrum in Figure \ref{f:DuffingcalibrationAmp}. The amplitude
of the mechanical displacement is given by

\begin{equation}
x=\frac{S_{II}^{\mathrm{meas}}(\Om)}{S_{II}^{\mathrm{meas}}(\Omod)}(\Om\phi_{0})\frac{\xzp}{g_{0}}
\end{equation}
From these measurements, the critical displacement $x_{c}=2.2\,\mathrm{nm}$
and the Duffing parameter $\beta=1.6\times10^{12}\,\mathrm{N/m^{3}}$
can be calculated\cite{Nayfeh1979}. 

\begin{figure}[htbp]
\centering \includegraphics[width=0.4\linewidth]{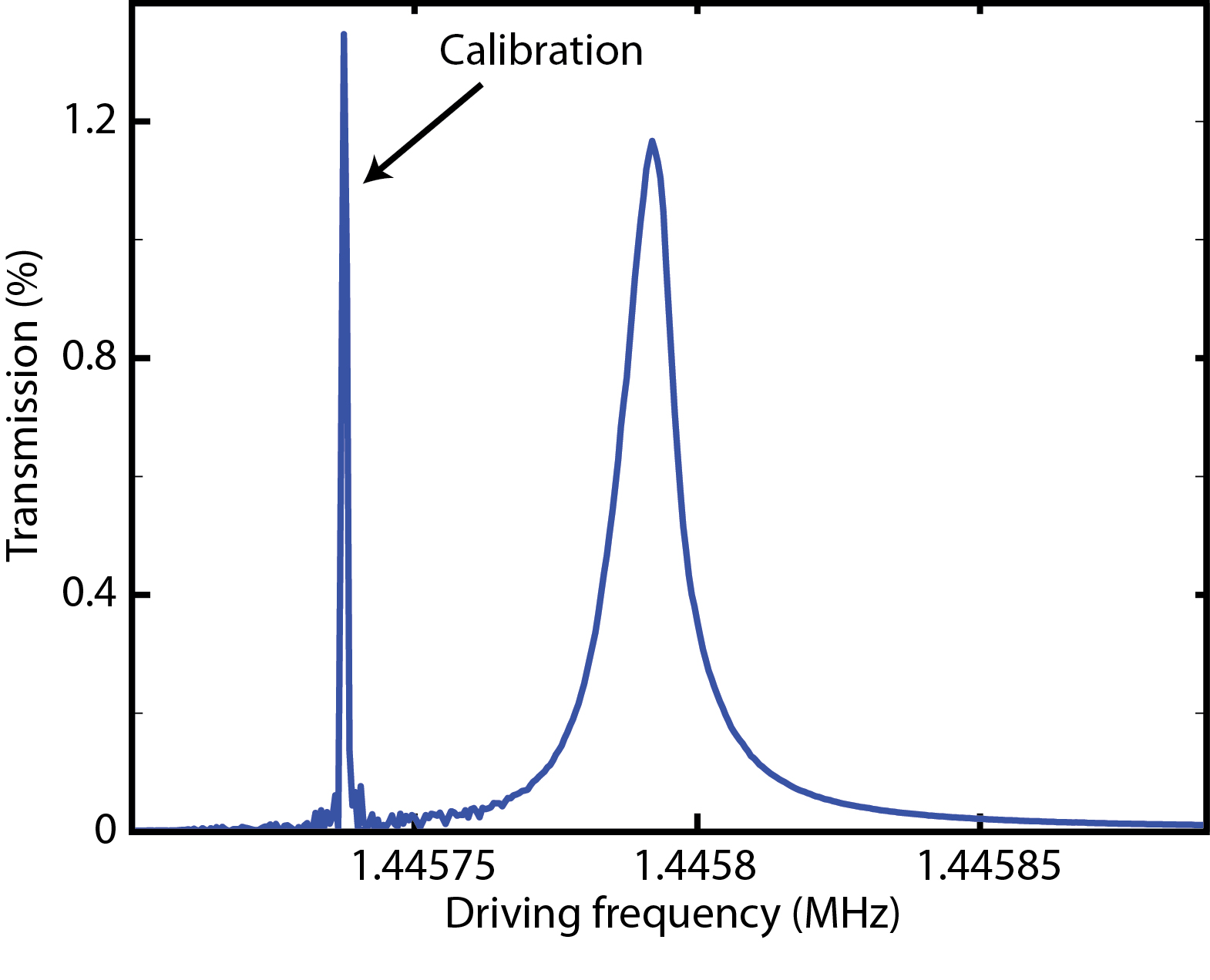}\caption{Homodyne readout spectrum of the driven mechanical oscillation and
the frequency modulation calibration. }

\label{f:DuffingcalibrationAmp} 
\end{figure}

\section{Theory of Slowing microwave}

The presence of the pump beam does not only induce a strong modification
of the transmission of the probe field, and at the same time leads
to a fast variation of the phase of the transmitted probe field across
the transmission window 

\begin{equation}
\phi=\mathrm{arg}(t_{\mathrm{p}})
\end{equation}
This can lead to significant group delays \cite{Schliesser2010,Weis2010,Safavi-Naeini2011,Jiang2011a},
analog to that achieved with the electromagnetically induced transparancy
in atomic \cite{Phillips2001,Liu2001} and in solid state media \cite{Longdell2005}.
\begin{equation}
\tau_{\mathrm{g}}=\frac{\partial\phi}{\partial\Omega}
\end{equation}
In the case of red-detuned pumping$\,$($\bar{\Delta}=-\Om$) and
$\Omega=\Om$, the group delay is approximated as, 

\begin{equation}
\tau_{\textrm{g}}=\frac{2\eta_{\mathrm{c}}C}{(1+C-\eta)(1+C)}\Gm^{-1}
\end{equation}
assuming only the mechanical susceptibility contributes to the phase
of the transmission spectrum$\,$($\Geff\ll\kappa$). Here $C=(2g_{0}\bar{a})^{2}/(\kappa\Gamma_{\mathrm{m}})$
denotes the optomechanical cooperativity parameter, and $\eta_{c}$
is cavity coupling parameter defined as the ratio of the cavity coupling
rate to the cavity damping rate$\,$($\eta_{\mathrm{c}}=\kappa_{\mathrm{ex}}/\kappa)$.
The maximum group delay is reached when 
\begin{equation}
\frac{\partial}{\partial C}\tau_{\textrm{g}}\equiv0
\end{equation}
The maximum group delay is given by 
\begin{equation}
\tau_{\mathrm{g}}^{\mathrm{max}}=2(1-\sqrt{1-\eta_{\mathrm{c}}})^{2}/\eta_{\mathrm{c}}\Gm
\end{equation}
when the cooperativity approaches $C=\sqrt{1-\eta_{\mathrm{c}}}$.

In the case with a slightly detuned probe tone, $\delta=\Omega-\Om$,
a good approximation of the group delay is given by 
\begin{equation}
\tau_{\mathrm{g}}\approx-\frac{\frac{2\eta_{\mathrm{c}}}{1-\eta_{\mathrm{c}}}C\Gamma_{\mathrm{m}}}{4\delta^{2}+(1+C)^{2}\Gamma_{\mathrm{m}}^{2}}\cdot\frac{4\delta^{2}-\frac{(1-\eta_{\mathrm{c}}+C)(1+C)}{1-\eta_{\mathrm{c}}}\Gamma_{\mathrm{m}}^{2}}{4\delta^{2}+\left(\frac{1-\eta_{\mathrm{c}}+C}{1-\eta_{\mathrm{c}}}\right)^{2}\Gamma_{\mathrm{m}}^{2}}
\end{equation}
This equation allows the group delay to be negative, i.e. the advancing
of the microwave waveforms, as long as the condition of $\vert\delta\vert>\Gm/2$
is met. 

The pulse delay is limited by the bandwidth of the transparency window$\,$($\Delta\nu\sim\Geff/2\pi$
), as the medium becomes increasingly opaque at frequencies outside
the transparency window, resulting in spreading of the pulse. In order
to preserve the pulse, the Gaussian pulse bandwidth should be smaller
than the bandwidth of the transmission window$\,$($\mathrm{BW_{p}<\Delta\nu}$).
In the main test, we study the group delay of a pulse with a full-width-half-maximum$\,$(FWHM)
duration of 83$\,$ms, which satisfies this condition. To study the
pulse distortion, a Gaussian pulse of 17$\,$ms FWHM is sent to to
cavity as a probe, and readout with a spectrum analyzer. As shown
in Figure$\,$\ref{f:PulseDelayDistortion}, the distortion reaches
its maximum at the largest group delay.

\begin{figure}[htbp]
\centering \includegraphics[width=0.5\linewidth]{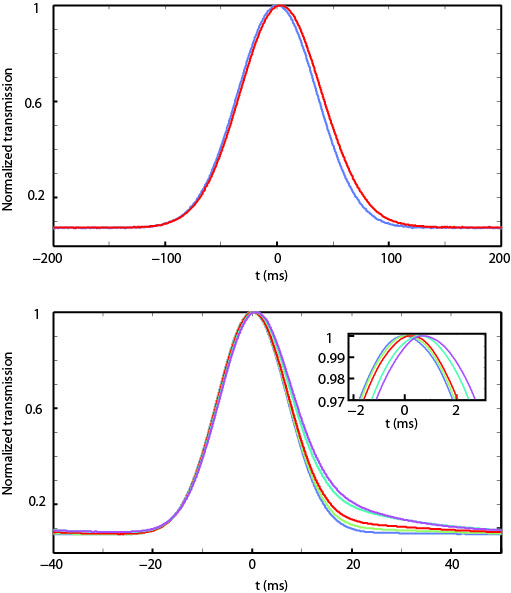}
\caption{Pulse delays and distortions. a) The non-delayed$\,$(blue) and delayed$\,$(red)
Gaussian pulses with FWHM of 83ms. b) The non-delayed and delayed
pulses with FWHM of 17ms. }

\label{f:PulseDelayDistortion} 
\end{figure}

\section{Theory of dynamical switching of optomechanically induced transparency}

\begin{spacing}{0.5}
We study the switching dynamics in a frame rotating at the pump frequency
$\opump$. We have the following set of ansatz
\begin{equation}
\delta a(t)=A^{-}(t)e^{-i\Omega t}+A^{+}(t)e^{+i\Omega t}
\end{equation}
\begin{equation}
\delta a^{*}(t)=\left(A^{+}(t)\right)^{*}e^{-i\Omega t}+\left(A^{-}(t)\right){}^{*}e^{+i\Omega t}
\end{equation}
\begin{equation}
\delta x(t)=X(t)e^{-i\Omega t}+X(t)^{*}e^{+i\Omega t}
\end{equation}
The intra-cavity field $\bar{a}(t)$ has a fast converge to its steady-state
value $\bar{a}$$\,$(either $\bar{a}_{\mathrm{on}}$ or $\bar{a}_{\mathrm{off}}$)
at a rate of $\kappa/2$. Applying the set of ansetz to equation$\,$(\ref{eq:eom1})(\ref{eq:eom2}),
and sort by rotation terms, the relevant equations with $e^{-i\Omega t}$
are

\[
\frac{d}{dt}A^{-}(t)=\left(i(\Omega+\bar{\Delta})-\frac{\kappa}{2}\right)A^{-}(t)-iG\bar{a}X(t)+\sqrt{\eta_{\mathrm{c}}\kappa/2}s_{\mathrm{P}}
\]

\end{spacing}

\begin{spacing}{0.20000000000000001}
\[
\frac{d}{dt}A^{+}(t)^{*}=\left(i(\Omega-\bar{\Delta})-\frac{\kappa}{2}\right)\left(A^{+}(t)\right)^{*}+iG\bar{a}X(t)^{*}
\]
\begin{equation}
\frac{d^{2}}{dt^{2}}X(t)+\frac{d}{dt}X(t)(\Gm-2i\Omega)+(\Om^{2}-\Omega^{2}-i\Gm\Omega)X(t)=-\meff^{-1}\hbar G\bar{a}(t)\left(A^{-}(t)+\left(A^{+}(t)\right)^{*}\right)\label{eq:X(T)eom}
\end{equation}
Note that we have assumed small excitation amplitudes so that the
Duffing nonlinearity is negligible. Due to the low mechanical damping
rate, we assume a the very slowing varying envelope of the displacement
of the mechanical oscillator, $\frac{d^{2}}{dt^{2}}X(t)\ll\Omega\frac{d}{dt}X(t)$,
$\Gm\frac{d}{dt}X(t)\ll\Omega\frac{d}{dt}X(t)$ with $\Omega\approx\Om$.
Further in the resolved-sideband regime, $A^{+}(t)\approx0$. The
cavity field amplitude $A(t)$ varies slowly as well, and therefore
we have $\vert\frac{d}{dt}A(t)\vert\ll\vert\frac{\kappa}{2}A(t)\vert$.
The set of equations$\,($\ref{eq:X(T)eom}) can then be simplified
as 
\end{spacing}

\begin{spacing}{0.5}
\[
0=i\left(\Omega+\bar{\Delta}-\frac{\kappa}{2}\right)\left(A^{-}(t)\right)^{*}-iG\bar{a}X(t)+\sqrt{\eta_{\mathrm{c}}\kappa/2}s_{\mathrm{P}}
\]

\end{spacing}

\begin{spacing}{0.20000000000000001}
\[
A^{+}(t)=0
\]

\end{spacing}

\begin{spacing}{0.5}
\begin{equation}
\frac{d}{dt}X(t)(-2i\Omega)+(\Om^{2}-\Omega^{2}-i\Gm\Omega)X(t)=-\meff^{-1}\hbar G\bar{a}A(t)\label{eq:Xtime}
\end{equation}
the solution of $A(t)\approx A^{-}(t)$ with detunning $-\bar{\Delta}=\Omega=\Om$
is approximately given by
\end{spacing}

\begin{equation}
A(t)=\frac{-iGX(t)\bar{a}+\sqrt{\eta_{\mathrm{c}}\kappa/2}s_{\mathrm{P}}}{\kappa/2}
\end{equation}
Note the temporal dynamics of the amplitude $A(t)$ is only determined
by the temporal dynamics of the mechanical displacement amplitude
$X(t)$. Using this solution in the equation of motion for $X(t)$,
we have

\begin{equation}
\frac{d}{dt}X(t)=-\left(\frac{\Gm}{2}+\frac{2\bar{a}^{2}G^{2}}{\kappa}\xzp^{2}\right)X(t)-iG\bar{a}\xzp^{2}\left(\frac{\sqrt{\eta_{\mathrm{c}}\kappa/2}s_{\mathrm{P}}}{\kappa/2}\right)\label{eq:Xt_solution}
\end{equation}

\subsection*{Switch-on dynamics}

Without loss of generality, we assume that at the time $t=0$ the
pump tone is injected into the cavity $P_{\mathrm{in}}(t)=P_{\mathrm{in}}[\mathrm{sgn(t)+1}]$
and the initial mechanical oscillation amplitude $X(0)=0$, the solution
of equation$\,$(\ref{eq:Xt_solution}) has the form
\begin{equation}
X(t)=X_{ss}\left(1-e^{-\frac{\Geff}{2}t}\right)
\end{equation}
where $X_{ss}=X(t\rightarrow\infty)=-iG\bar{a}_{\mathrm{on}}\xzp^{2}\left(\frac{\sqrt{\eta_{\mathrm{c}}\kappa/2}s_{\mathrm{P}}}{\kappa/2}\right)$
is its steady-state amplitude. Physically this means that $X(t)$
rings up to its steady-state amplitude $X$ at a rate $\Geff=\frac{\Gm}{2}+\frac{2\bar{a}_{\mathrm{on}}^{2}G^{2}}{\kappa}\xzp^{2}$.
The solution of $A(t)$ is then given by 
\begin{equation}
A(t)=\frac{-iGX_{ss}\left(1-e^{-\frac{\Geff}{2}t}\right)\bar{a}_{\mathrm{on}}+\sqrt{\eta_{\mathrm{c}}\kappa/2}s_{P}}{\kappa/2}
\end{equation}
having the initial amplitude $A(0)=\frac{\sqrt{\eta_{\mathrm{c}}\kappa/2}s_{\mathrm{P}}}{\kappa/2}$,
and the steady state amplitude $A_{ss}=A(t\rightarrow\infty)=\frac{-iGX_{ss}\bar{a}_{\mathrm{on}}+\sqrt{\eta_{\mathrm{c}}\kappa/2}s_{\mathrm{P}}}{\kappa/2}$.
This implies that theintra-cavity amplitude $A(t)$ rings down at
a rate close to $\Geff/2$ and reaches its stead state amplitude $A_{ss}$.
The readout of the transmitted field $t_{\mathrm{p}}=1-\sqrt{\eta_{\mathrm{c}}\kappa/2}A(t)/s_{\mathrm{P}}$,
therefore, rings up at a rate close to $\Geff/2$, as shown in Figure
\ref{f:RingUp}.

\subsection*{Switch-off dynamics}

Moreover, when the pump field is switched off, the intra-cavity photon
from the pump $\bar{a}(t)$ decays to $\bar{a}_{\mathrm{off}}\approx0$
at a fast rate of $\kappa/2$, resulting in a fast increase of the
intra-cavity field to $A(t)\mid_{\mathrm{pump-off}}=\frac{\sqrt{\eta_{\mathrm{c}}\kappa/2}s_{\mathrm{P}}}{\kappa/2}$,
corresponding to a fast decay of transmission amplitude $t_{\mathrm{p}}$
shown in Fig.\ref{f:RingUp}a) at $t=150\,\mathrm{ms}$. Due to the
absence of the pump tone$\,$($\bar{a}$=0), and further with $X(t)\mid_{\mathrm{pump-off}}=X_{ss}$,
the mechanical oscillation amplitude$\,$(\ref{eq:Xt_solution}) decays,
in this case, as 

\begin{equation}
X(t)=X_{ss}e^{-\frac{\Gm}{2}t}
\end{equation}
A qualitative measure of this decay is given by the Raman-scattered
field amplitude $B_{0}(t)$ at the blue sideband of the cavity$\,$(i.e.
$\omega_{c}+\Om$), to which the mechanical oscillation scatters part
of the probe light even in absence of the pump. As $B_{0}(t)$ is
proportional to $\left(-\frac{\sqrt{\eta_{\mathrm{c}}\kappa/2}s_{\mathrm{P}}}{\kappa/2}\right)X(t)$,
the scattered field $B_{0}(t)$ rings down at a rate of $\Gm/2$,
as shown in Figure \ref{f:RingUp} b). The strong initial rise is
associated with the fast rise of the intra-cavity probe amplitude
to $A(t)\mid_{\mathrm{pump-off}}$ as the pump field is switched off. 

\begin{figure}[htbp]
\centering \includegraphics[width=0.8\linewidth]{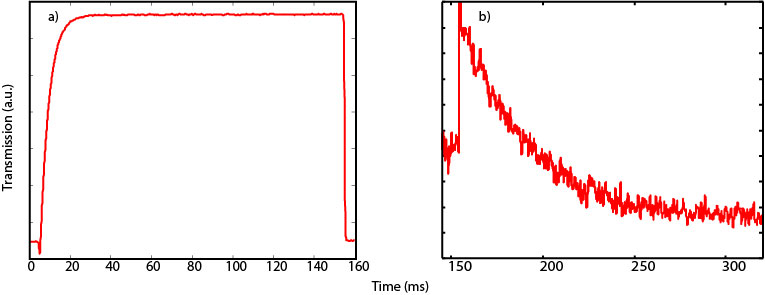}\caption{Probe field transmission $t_{\mathrm{p}}$. a) Probe on cavity resonance$\,$($\ocavity$).
The probe transmission rings up at the rate of $\Geff$/2 upon switching
on the pump tone. The probe transmission immediately drops at $T=$150$\,$ms
as pump tone switches off. b) Probe blue-detuned to the cavity ($\ocavity+\Om$).
The probe transmission increases at a fast rate$\,$($\kappa/2$)
upon switching off of the pump tone, and follows by a decay at a rate
of $\Gm/2$.}

\label{f:RingUp} 
\end{figure}

\end{widetext}

\end{document}